\documentclass[twocolumn,showpacs,preprintnumbers,amsmath,amssymb,aps,prl]{revtex4}
\usepackage{graphicx}
\def\bea{\begin{eqnarray}}
\def\eea{\end{eqnarray}}
\def\be{\begin{equation}}
\def\ee{\end{equation}}
\def\z{z^L}
\def\S{\mbox{\bf S}}
\def\et{{\it et al.}}
\begin{document}
\author{A.\ L\"auchli$^1$, G.\ Schmid$^1$ and M.\ Troyer$^{1,2}$}
\affiliation{$^1$\ Institut f\"ur Theoretische Physik, ETH H\"onggerberg, CH-8093
Z\"urich,Switzerland\\
$^2$\ Computation Laboratory, ETH Z\"urich, CH-8092 Z\"urich, Switzerland}
\date{\today}
\title{Phase diagram of a spin ladder with cyclic four-spin exchange}
\begin{abstract}
We present the phase diagram of the $S=1/2$ Heisenberg model on the two leg ladder with cyclic four
spin exchange, determined by a combination of Exact Diagonalization and Density Matrix
Renormalization Group techniques. We find six different phases and regimes: the rung singlet phase,
a ferromagnetic phase, two  symmetry broken phases with staggered dimers and staggered scalar
chiralities, and a gapped region with dominant vector chirality or collinear spin correlations. We
localize the phase transitions and investigate their nature.
\end{abstract}
\maketitle Frustrated interactions in quantum spin systems give rise to new and exotic phases but
are little understood because of the inherent difficulties with competing interactions. A special
type of frustration due to cyclic exchange interactions was recently found to be important in the
spin ladder material La$_x$Ca$_{14-x}$Cu$_{24}$O$_{41}$~\cite{TelephoneCompound} and in cuprate
antiferromagnets such as La$_2$CuO$_4$~\cite{Coldea214}. Similar multiple spin exchange
interactions are known to be relevant for the nuclear magnetism of $^3$He \cite{LhuillierTriangle}
and for the spin structure of a Wigner crystal \cite{Klaus}.
\begin{figure}[b]
\begin{center}
\includegraphics[width=0.8\linewidth]{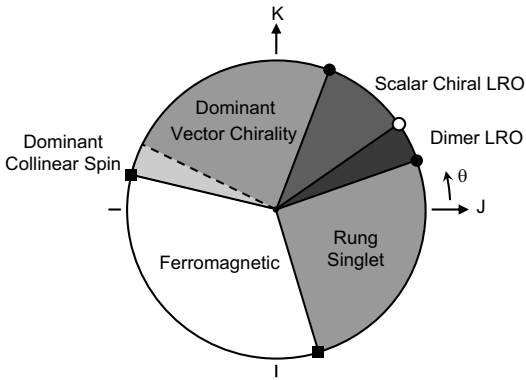}
\caption{ Phase diagram of the two leg ladder with cyclic four spin interaction. Squares denote
first order phase transitions, and the full circles indicate second order phase transitions. The
nature of the phase transition marked with the empty circle is presently unknown. The dashed line
indicates a crossover line without a phase transition. }
\label{PhaseDiagram}
\end{center}
\end{figure}

In contrast to these systems on the triangular lattice, where already the nearest neighbor
antiferromagnetic spin exchange is frustrating, in the bipartite two-dimensional (2D) square
lattice model, relevant for the cuprates, the frustration only enters through the cyclic exchange
term. It has been proposed that on the square lattice exotic magnetic phases with fractionalized
excitations could exist \cite{SenthilFisher}.
However, experience gained from quantum dimer models, showing a dimer liquid phase on a triangular
lattice\cite{Moessner} but not on the square lattice \cite{RK} indicates that on a bipartite
lattice ordered phases are preferred and exotic spin liquid phases hard to realize.

In our numerical studies of a cyclic exchange model we consider a ladder geometry, which is the
simplest system on which cyclic exchange is possible. Consisting of two coupled chains, ladders
already exhibit behavior reminiscent of two-dimensional systems and can  be investigated accurately
using the density matrix renormalization group (DMRG) method \cite{SRWDmrg}.  While there exist a
number of interesting ladder materials the aim of our study is not the discussion of these
materials (which all seem to be in the Haldane phase). Rather, the results obtained give insight
also into the properties of the 2D system.

Previous numerical studies on ladders \cite{LadderBrehmer,LadderHonda,CftEd,Nunner,MuellerMikeska}
were restricted to small cyclic exchange terms, only up to the critical point where the spin gap
closes and a new phase appears. We go beyond this weakly frustrated regime and present the phase
diagram at all strengths and signs of the bilinear and cyclic four spin exchange. Our SU(2)
invariant Hamiltonian on a $S=1/2$ two leg ladder is defined as follows: \bea H&=&J_{\perp}\
\sum_{x} \mbox{{\bf
    S}}(x,1)\cdot\mbox{\bf{S}}(x,2)\nonumber \\
&&+ J_{\parallel}\ \sum_{x,y} \mbox{{\bf
    S}}(x,y)\cdot\mbox{\bf{S}}(x+1,y)\label{Hamiltonian} \\
&&+ K\ \sum_{x} \left[P_{\square}^{\mbox{ }}(x,x+1)+P_{\square}^
  {-1}(x,x+1)\right],\nonumber
\eea where $J_\perp$ $(J_\parallel)$ are the bilinear exchange constants on the rungs (along the
legs) of the ladder and $K$ denotes the coupling of the cyclic four spin permutation operators.
This operator can be decomposed in terms of spin operators involving bilinear and biquadratic terms
\cite{spinrep}. We set $J_\perp=J_\parallel=J$ in the following and parameterize the couplings as
$J=\cos(\theta)$ and $K=\sin(\theta)$. The energy scale $\sqrt{J^2+K^2}$ is set to one.

The numerical algorithms we employ are exact diagonalization (ED) of finite systems up to $2 \times
16$ sites with periodic boundary conditions (PBC), and the finite size version of the DMRG
algorithm \cite{SRWDmrg} on systems with up to $2 \times 200$ sites, keeping up to 1000 states and
using appropriate open boundary conditions (OBC) \cite{NoteBoundaryDMRG}. We carefully checked the
convergence of our results with respect to the lattice size and the number of states kept.

The phase diagram in Fig. \ref{PhaseDiagram} summarizes our results. We now proceed to characterize
the phases based on their order parameter or their slowest decaying correlation function (termed
{\it dominant} correlation function) and then discuss the phase transitions, crossovers and
universality classes.

{\it Rung Singlet Phase -- } We start the discussion of the phase diagram at $\theta=0$, i.e the
spin ladder with only antiferromagnetic bilinear couplings. Here the groundstate is unique and well
approximated by the product of local rung singlets, hence the name {\it rung singlet phase}. All
excitations are gapped and the correlation functions decay exponentially. The dominant correlations
are the spin-spin correlations. A small positive $K$ is sufficient to close the gap and to drive
the system into a new phase \cite{LadderBrehmer,LadderHonda,CftEd,Nunner,MuellerMikeska}. We find
that a negative $K$ has a less pronounced effect. All correlation functions decay even faster and
at $\theta=-0.40\ \pi$ $(J/|K|=0.30)$ we locate a first order transition to the ferromagnetic
phase.

\begin{figure}
  \includegraphics[width=5cm]{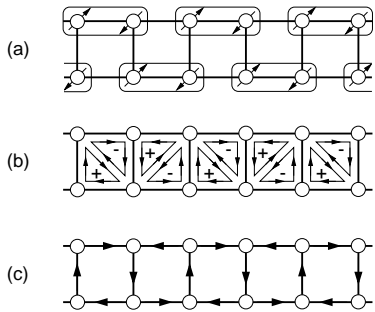}
  \caption{
    (a) One of the two degenerate groundstates in the long range ordered
    staggered dimer phase. The spins are paired in singlets.
    (b) Spatial structure of the scalar chirality order parameter in one
    of the two symmetry breaking groundstates. The oriented triangle
    (the sign) denotes the
    orientation of the triple product (the sign of the order parameter).
    (c) Correlation pattern in the dominant vector chirality region.
    Oriented bonds stand for the cross product of two spins.
    Correlations between two such bonds are positive.
    }
  \label{orderpatterns}
\end{figure}

\begin{figure}
  \includegraphics[width=0.8\linewidth,angle=0]{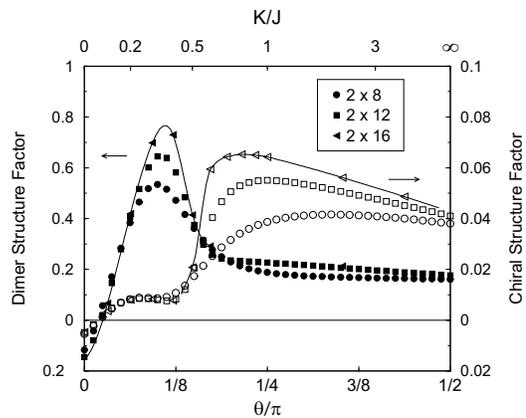}
  \caption{ED calculations of the staggered dimer structure factor
    \protect{\cite{DimerStructureFactor}} (filled symbols)
    and the scalar chirality structure factor (open symbols) in the
    ground state. Lines are a guide to the eye. The increase of the
    structure factor with
    system size is an indication for long range order.
    }
  \label{DimerChiralED}
\end{figure}

\begin{figure}
  \includegraphics[width=0.7\linewidth,angle=0]{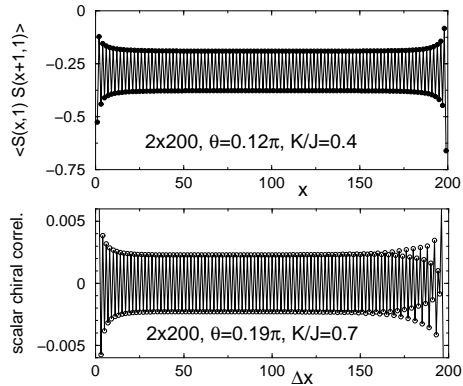}
  \caption{Upper panel: local $\S(x,y)\cdot\S(x+1,y)$ expectation value
    on one of the two legs in the dimer LRO phase at $\theta=0.12\ \pi$.
    The OBC render a direct measurement of this order parameter possible.
    Lower panel:
    long distance behavior of the scalar chirality correlations between
    equally oriented triangles in the scalar chiral LRO phase at
    $\theta=0.19\ \pi$.
    }
  \label{DimerChiralDMRG}
\end{figure}

{\it Staggered Dimer Phase -- } Between $\theta=0.07\pm0.01\ \pi$ $(K/J=0.23\pm0.03)$ and
$\theta\approx 0.15\ \pi$ $(K/J\approx 0.5)$ the system is in a valence bond crystal phase with a
staggered dimer pattern and a finite gap to triplet excitations. The order parameter is: \be
\langle\S(x-1,y)\cdot\S(x,y)-\S(x,y)\cdot\S(x+1,y)\rangle \ee with a spatial structure as shown in
Fig.~\ref{orderpatterns}~(a). This phase breaks the translation symmetry and has a twofold
degenerate groundstate. Since this is a broken {\it discrete} symmetry, long range order (LRO) can
exist even in one dimension. This order is seen in ED calculations (full symbols in
Fig.~\ref{DimerChiralED}) and in DMRG results (upper panel of Fig.~\ref{DimerChiralDMRG}). Since
the boundary conditions \cite{NoteBoundaryDMRG} in the DMRG calculations select one of the two
possible groundstates, the order parameter can be measured directly and shows convincing evidence
for long range staggered dimer order. Finally, the doubling of the unit cell in the symmetry broken
phase manifests itself in degenerate singlets at momenta $(0,0)$ and $(\pi,\pi)$ in the infinite
system. We have confirmed the existence of two nearly degenerate states at these momenta in ED,
with a small finite size splitting.


{\it Scalar Chirality Phase -- } For $\theta$ larger than $\approx 0.15\ \pi$ the dimerization
vanishes and we find a gapped phase with LRO in the staggered scalar chirality. The order parameter
is: \be \langle \S(x,1)\cdot \left[ \S(x,2) \times \S(x+1,1)\right] \rangle \ee and has a spatial
modulation with wave vector $(\pi,\pi)$ (see Fig.~\ref{orderpatterns}~(b) for a pictorial
representation). This order parameter breaks spatial symmetries and time reversal symmetry, but not
SU(2). LRO in this unexpected phase is seen as before in i) ED calculations of the corresponding
structure factor (Fig.~\ref{DimerChiralED}, open symbols), ii) DMRG calculations of the order
parameter correlations, converging to a finite value at large distances (lower panel of
Fig~\ref{DimerChiralDMRG}), iii) the existence of a $(\pi,\pi)$ singlet which is energetically
close to the groundstate. The discrete symmetry breaking in this phase suggests a finite triplet
gap. We find that in both the dimerized and the scalar chirality LRO phases the triplet gap is
finite but small [$\Delta(S=1)\lesssim 0.1$].

\begin{figure}
  \includegraphics[width=0.9\linewidth,angle=0]{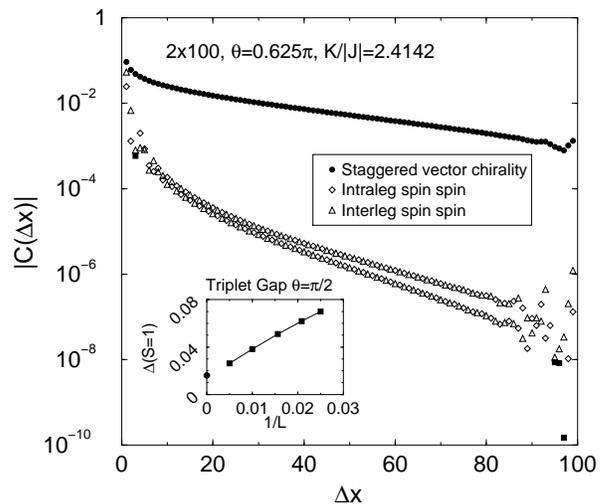}
  \caption{Semi-log plot of the long distance behavior of the
    staggered vector chirality correlations on the rungs and
    the spin-spin correlations for $\theta=5\pi/8$.
    The spin correlations decay much faster than the vector chirality
    correlations.
    The open (filled) symbols for the spin correlations denote negative
    (positive) correlations.
    Inset: finite size scaling of the spin gap at $\theta = \pi/2$.
    }
  \label{CorrsNematic}
\end{figure}

{\it Dominant Vector Chirality Region -- } At $\theta=0.39\pm 0.01\ \pi$ $(K/J=2.8\pm0.3)$ we
locate a second order phase transition to a short range ordered phase with a unique groundstate and
a fully gapped excitation spectrum similar to the rung singlet phase. In contrast to the rung
singlet phase the dominant groundstate correlations are not the spin-spin correlations, but
correlations of the following vector chirality order parameters: \be \S(x,y)\times \S(x+1,y),\quad
\S(x,y)\times \S(x,y+1) \ee in staggered circulation arrangement [Fig.~\ref{orderpatterns}~(c)].
This vector chirality is also called {\it twist} or {\it helicity} and can be regarded as a local
spin current operator for bilinear Heisenberg Hamiltonians. Correlations are strong between bonds
on rungs or legs, but diagonal bonds  are very weakly correlated. The vector chirality and the spin
correlations shown in Fig.~\ref{CorrsNematic} for DMRG calculations at $\theta=5\pi/8$ clearly
demonstrate that the vector chirality correlations decay much slower $($correlation length
$\xi\approx 30)$ than the spin-spin correlations $(\xi\approx 12)$.
The existence of a small but finite gap in this region is confirmed by a finite size scaling of the
triplet gap for the case of pure $K$ $(\theta=\pi/2)$ in the inset of Fig.~\ref{CorrsNematic} (DMRG
results). An extrapolation in $1/L$ ($L\le 200$) yields a lower bound on the infinite system gap:
$\Delta(S=1) \ge 0.016 K$. Computations of dynamical spin and vector chirality structure factors in
this region reveal another striking difference compared to the Rung Singlet phase: the lowest
triplet excitation, with wavevector $(\pi,\pi)$, is not a magnon (i.e. a spin flip excitation) but
rather of the {\it vector chirality type}. It exhausts a large fraction of the spectral weight in
the vector chirality structure factor. The presence of vector chirality correlations for pure $K$
can be anticipated at the classical level. The groundstate has a four sublattice structure where
nearest neighbor spins are orthogonal to each other \cite{Chubukovfourspin} and therefore maximize
the vector chirality. Our results now suggest that the transition from the classical to the $S=1/2$
quantum case leads to short range order in both the spin and the vector-chirality correlations,
with the latter becoming dominant. At $\theta \approx 0.85\ \pi$ $(K/|J|\approx 0.5)$ there is a
crossover region to dominant collinear spin correlations.

{\it Dominant Collinear Spin Region -- } In the proximity of the ferromagnetic phase boundary a
short range ordered region characterized by collinear $(0,\pi)$ spin-spin correlations is observed,
where spins on the same leg  (on different legs) exhibit ferromagnetic (antiferromagnetic)
correlations. The system has a unique groundstate and a fully gapped spectrum.

{\it Ferromagnetic Phase -- } The last phase is the fully polarized ferromagnetic phase, located
between two first order transitions at $\theta=0.94\ \pi \ (K/|J|=0.19)$ and  $\theta = -0.40\ \pi\
(J/|K|=0.30)$. This phase extends beyond the rigorous bounds  $-\pi \le \theta \le -\pi/2$, inside
which the ferromagnetic state minimizes the energy on each plaquette separately.

{\it Phase Transitions and Universality Classes -- } We apply  the method of the Lieb-Schulz-Mattis
twist operators to our system in order to precisely locate phases transitions and to discuss
universality classes by considering the following quantity \cite{NakamuraTodo1}: \be
\z=\langle\mbox{GS}| \exp\left[{\rm i}\frac{2\pi}{L}\sum_{x=1}^L x
  \tilde{S}^z(x)\right]|\mbox{GS}\rangle,
\label{zdefinition} \ee where $\tilde{S}^z(x)=\mbox{\bf S}^z(x,1)+\mbox{\bf S}^z(x,2)$ and $L$ is
the system length.
\begin{figure}
  \includegraphics[width=0.7\linewidth]{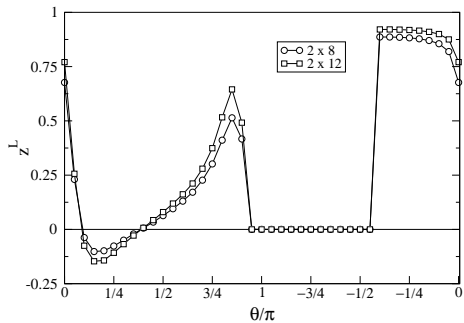}
  \caption{
    ED results of $\z$ (definition in the text)
    as a function of $\theta$. The infinite system values of this
    operator are $\pm 1$ or $0$. For clarity not all system sizes are shown.
    }
  \label{thetaz}
\end{figure}
The infinite system  values of this quantity converge to $\pm 1$ or $0$. In the Rung Singlet phase
$\z$ tends to $+1$ \cite{NakamuraTodo1}. The transition at $\theta=0.07\ \pi$ is signaled by a sign
change in $\z$ and can therefore be determined quite accurately. Our scenario of a transition from
the Haldane universality class of the Rung Singlet phase to the dimerized phase is reminiscent of
the Babudjian-Takhtajan (BT) type transition in the $S{=}1$ bilinear-biquadratic chain
\cite{Babudjian-Takhtajan} and this supports results obtained in
\cite{Tsvelik,CftEd,MuellerMikeska}. The phase transition between the dimerized phase and the
scalar chirality LRO phase ($\theta\approx0.15\ \pi$) has no visible signature in the behavior of
$\z$. The rapid change of the groundstate correlations at this transition make it difficult to
discern between a first or second order phase transition. The second order phase transition at
$\theta=0.39\ \pi$ is detected again by a sign change in $\z$. The universality class of this phase
transition remains for further study. There is no evidence for a phase transition between the
dominant vector chirality and the dominant collinear spin regions but we see a smooth crossover
with stable gaps instead. The ferromagnetic region finally shows vanishing $\z$ and the transitions
to it are of first order.

{\it Conclusions -- } The two leg ladder with cyclic four spin exchange reveals a very rich phase
diagram. Besides determining the domain of stability of conventional phases such as the rung
singlet phase, the collinear spin region and the ferromagnetic phase we have established a
dimerized phase and chiral phases. Dimerized phases are common in frustrated spin chains but do not
appear as generic phases of the diagonally frustrated two leg ladder \cite{RogerPhaseDiagram}.
However several authors have shown the existence of dimerized phases in ladder models including
biquadratic terms \cite{Tsvelik,biquadraticladder} and earlier studies  of the present Hamiltonian
\cite{LadderBrehmer,CftEd,MuellerMikeska} conjectured a dimerized phase for large $K/J$ -- which we
have now confirmed.

Unexpectedly we also found two chiral regions, one with long range order in the staggered scalar
chirality and a second with dominant vector chirality correlations. Uniform scalar chirality phases
have a long history in anyon superconductivity \cite{Wen} and were discussed in the context of
frustrated spin models \cite{J1J2chiral} but were not seen or conjectured in the context of ladder
models.

Our results on ladders give insights also into the phase diagrams of cyclic exchange models on the
square lattice \cite{Chubukovfourspin}, since long range ordered phases on the ladder can be
further stabilized when they are coupled to form 2D planes. We conjecture that the dominant vector
chirality region leads to a $T=0$ long range ordered vector chirality phase. This state can be
regarded as a {\it spin nematic} which has long range order in the twist correlations, but no
magnetic moment \cite{ChandraColeman}. The existence of such a phase has been conjectured, but a
realization has not been observed in a microscopic model up to now. In fermionic terms this state
is related to triplet $d$-density wave states with staggered spin currents \cite{ddw}.

We would like to thank I.~Affleck, M.P.A.~Fisher, A.~Kampf, C.~Lhuillier, M.~Sigrist, R.R.P.~Singh,
S.~Todo, and S.~Wessel for very fruitful discussions. G.~S. and M.~T. acknowledge support from
Swiss National Science Foundation. We would like to thank S.\ R.\ White for help with the DMRG
code. 

\begin{references}
\bibitem{spinrep}
  The representation of $P_{\square}+P_{\square}^{-1}$
  in terms of spin operators can be found in
  ref. \cite{MuellerMikeska} for example.
\bibitem{NoteBoundaryDMRG}
  We used standard open boundary conditions in all the regions but the
  dimerized and the scalar chirality LRO
  phase. In these cases we attached one more $(S{=}1/2)$ spin at each
  end of the ladder in the same way as
  Ref. \cite{RogerPhaseDiagram}.
\bibitem{DimerStructureFactor}
  We define the dimer structure factor in this context as the
  staggered sum of all leg dimer dimer correlations where the two bonds
  do not overlap.
\end{references}
\end{document}